\documentclass[aps,prd,onecolumn,showkeys,amssymb]{revtex4}
\usepackage{float}
\usepackage{amssymb}
\usepackage{graphicx}
\usepackage{amsmath}
\usepackage{hyperref}
\usepackage{subcaption}
\usepackage{adjustbox}
\usepackage{dcolumn}
\usepackage{xcolor}

\def\be{\begin{equation}}
\def\ee{\end{equation}}
\def\beq{\begin{eqnarray}}
\def\eeq{\end{eqnarray}}

\begin{document}

\title{ Isotropic compact stars in  4D Einstein-Gauss-Bonnet gravity }
 
\author{Sudan Hansraj$^{1}$, Ayan Banerjee $^{1}$,   Lushen Moodly$^{1}$ and M. K. Jasim$^{2}$  }

\affiliation{$^{1}$ Astrophysics and Cosmology Research Unit, University of KwaZulu Natal, Private Bag X54001, Durban 4000,
South Africa,\\
$^4$Department of Mathematical and Physical Sciences, College of Arts and Science, University of Nizwa, Nizwa, Sultanate of Oman}

\begin{abstract}

Recently it has been proposed that the Gauss-Bonnet coupling parameter of Lovelock gravity may  suitably be rescaled in order to admit physically viable models of celestial phenomena such that higher curvature effects are active in standard four dimensions as opposed to the usual higher dimensions.  We investigate the consequences of this modification in the context of stellar modelling. The evolution of  perfect fluid distributions  is governed by the pressure isotropy condition and through stipulation of one of the metric potentials complete models emerge from solutions of the master differential equation. New classes of exact solution with this approach have been reported. One particular model is analysed in detail and shown to comport with elementary physical requirements demanded of realistic compact stars suggesting that the  modified theory is not inconsistent with observations.

\end{abstract}


\maketitle
\flushbottom

\section{Introduction}\label{intro:Sec}

In a recent contribution Glavan and Lin  \cite{Glavan:2019inb} proposed that the Einstein--Gauss--Bonnet (EGB) gravity theory, which is well known to be quintessentially a higher dimensional theory with extra curvature terms that are dynamic in dimensions 5 and 6, may in fact be studied in the four dimensional context by a scaling of the Gauss-Bonnet coupling $\alpha$ by a factor of $\alpha/(D -4)$,  and taking the limit $D \to 4$. A major advantage of this procedure is that the Lovelock theorem demanding up to second order equations of motion is satisfied and Ostrogradsky instability is avoided. The mathematical consequences to this approach are mitigated by the   dimensional regularization process proposed by  \cite{Tomozawa:2011gp}. One application of this proposal was made by Glavan and Lin in generating a static spherically symmetric singularity--free black hole solution. It turns out that  this interesting scenario  is different from those of black hole solutions in higher dimensional EGB gravity. 

Notwithstanding the ostensible  positive features of the 4$D$ EGB  theory there are several criticisms against the proposal. At the outset, subjecting a discrete element such as spacetime dimension to a limiting process has evoked concern but the dimensional regularization technique endeavours to deal with this aspect \cite{Tomozawa:2011gp}.     Gurses {\it{et al}} \cite{Gurses:2020ofy,Ai:2020peo,Mahapatra:2020rds,Shu:2020cjw}  have discussed two independent lines of reasoning to show why such a theory is not palatable in a four dimensional setting in general. These authors do however agree that in spherically symmetric spacetimes there may be no objection and so in this article we endeavour to analyse the behaviour of such a theory in the context of static stellar modelling. In particular, our interest lies in determining whether this theory admits physically viable compact star models. 

The literature is replete with investigations of other configurations of matter fields in 4D EGB. For example,  rotating and non-rotating  black hole solutions and their physical properties   have received attention in 4D EGB theory \cite{Ghosh:2020syx,Konoplya:2020juj,Kumar:2020uyz,Kumar:2020xvu,Zhang:2020sjh,Liu:2020vkh}. In addition,  the strong/ weak gravitational lensing problem  by black holes  was considered by  \cite{Islam:2020xmy,Kumar:2020sag,Heydari-Fard:2020sib,Jin:2020emq}. The question of  geodesic  motion and the shadow of black holes was investigated by  \cite{Zeng:2020dco} while 
Hawking radiation was considered  by \cite{Zhang:2020qam,Konoplya:2020cbv}. Quasinormal modes  were discussed by  \cite{Churilova:2020aca,Mishra:2020gce} and 
wormhole and thin-shell wormhole solutions generated  by \cite{Jusufi:2020yus,Liu:2020yhu}.  For a number of other matter configurations that have been treated in the literature   see the recent works of  
\cite{Easson:2020mpq,Jusufi:2020qyw,Yang:2020jno,Ma:2020ufk,Dadhich:2020ukj,Samart:2020sxj}.  

Note also that there are other attempts at studying the effects of higher curvature EGB terms in a four dimensional setting by the inclusion of  a scalar field such as a dilaton. This is known   as Einstein dilaton Gauss-Bonnet (EdGB) gravity and  has been extensively considered by \cite{Boulware:1986dr,Kanti:1995vq,Kanti:1997br,Torii:1996yi,Nojiri:2005vv,Nojiri:2006je,Leith:2007bu,Bamba:2007ef,Niz:2007gq,Capozziello:2008gu,Ohta:2009pe,Maeda:2009uy,Guo:2009uk,Guo:2010jr,Chen:2009rv,Nojiri:2010wj}. The major drawback of this proposal is that it involves a distortion of the Einstein frame necessitating a conformal transformation of the frame \cite{Iihoshi:2011ew}. This procedure then generates extra couplings with matter and these are generally conveniently ignored \cite{bouldes2} when models are developed. Nevertheless numerous explorations of this proposal have been undertaken.  Early studies  on EdGB gravity focussed on the low-energy effective  action of the bosonic sector of heterotic string theory \cite{Kanti:1995vq,Pani:2009wy}.   Recently, BH solutions have been found for more general coupling functions \cite{Antoniou:2017acq}. A study of neutron stars in the EdGB gravity was conducted  in Ref. \cite{Pani:2011xm} including both the static and the slowly rotating cases. It is worth mentioning  that for certain fixed values of the coupling parameter no neutron star solutions exist above some critical 
central energy density as discussed in  \cite{Pani:2011xm}. In the context of these theories, the axial  quasinormal modes of neutron stars were examined in \cite{Blazquez-Salcedo:2015ets}, and 
additional neutron star solutions were constructed by  \cite{Doneva:2017duq} (see Ref. \cite{Blazquez-Salcedo:2018pxo} 
for further reviews).   In addition  traversable wormholes were numerically derived with a coupling function of exponential type with no need to resort to exotic matter in (3 + 1)-dimensions. In a similar vein some authors  \cite{Cuyubamba:2018jdl}  have shown  that the well known  `Kanti-Kleihaus-Kunz' wormhole is unstable against small perturbations for any values of its parameters. 

In this article we probe the novel 4$D$ approach through rescaling of the GB parameter. Before elucidating the  proposal we consider briefly the motivation to pursue extensions of the Einstein theory of gravity which, despite its  numerous experimental successes,  leave  many questions  unanswered. 
The current accelerated expansion of the universe has been confirmed by several independent observations, including supernovae Ia data \cite{Perlmutter}, Baryon Acoustic Oscillations \cite{Eisenstein} and cosmic microwave background radiation (CMB) \cite{Spergel} as studied by the WMAP. This raises many intriguing questions amongst them is the  missing mass problem  which was noted for the first time by Fritz Zwicky \cite{Zwicky}. The puzzle can be addressed either introducing some form of dark energy or assuming some kinds of modifications to the classical GR.  Several such ideas have been explored including higher derivative gravity (HDG) theories such as $f(R)$ gravity \cite{Starobinsky}.   In particular, HDG was motivated by the search  for singularity free black holes solutions, avoiding causality problems at the classical level and so on. An alternative modification of GR is to add higher curvature effects through polynomial invariants of the Riemann tensor, Ricci tensor and Ricci scalar without compromising diffeomorphism invariance or the Bianchi identies. In this regard, Lovelock gravity (LG) \cite{Lovelock,Lovelock:1972vz} is a  natural generalization of Einstein's general relativity. In this higher curvature effects are only dynamic for dimensions higher than 4. Additionally  Lovelock gravity preserves  conservation of energy momentum and are free from unique ghosts  when expanded on a flat space,   avoiding problems with unitarity \cite{Zumino:1985dp,Zwiebach:1985uq}. At the quadratic level the Lovelock action is called  Gauss–Bonnet (GB) action or the Lanczos action  \cite{Lanczos:1938sf}. This is the simplest nontrivial case to study higher curvature effects.  Interestingly, The GB action appears in the low energy effective action of heterotic string theory  \cite{Wiltshire:1985us,Boulware:1985wk,Wheeler:1986}.

Since GB terms are known to affect physics in dimensions higher than four it is interesting to study the viability of the 4D EGB version in the context of stellar modelling. Compact stars and their properties have been an active area of research  for many decades. The existence of a mass limit for  white dwarfs by Chandrasekhar \cite{Chandrasekhar} advanced the way to  understanding the  nature and structure of compact stars. In his work, it was shown that white dwarfs are compact stars in which the pressure support against collapse comes from the  quantum degeneracy of the electrons. In the context of GR there exists  a large number of works  studying the behaviour of static spherically symmetric objects composed of a perfect fluid. A comprehensive listing of such exact solutions for isotropic fluids may be found in \cite{Delgaty:1998uy}.  Exact solutions in five dimensional
EGB theory were generated in  \cite{Maharaj:2015gsd,Chilambwe:2015rra,Dadhich:2015rea,Hansraj:2016gmk} and references therein.   The objective of this paper is to find new exact interior models in  4$D$ EGB theory 
 with a spherical distribution of perfect fluid. Some attempts in this direction have been made by 
(see refs. \cite{Doneva:2020ped,Banerjee:2020stc,Banerjee:2020yhu}) where authors   aimed to address several shortcomings coming out in the study of compact objects at high densities.  Recently, Banerjee \emph{et al} \cite{Banerjee:2020dad} studied spherically symmetric static solutions in this scenario with interacting quark EoS. The effects of the coupling constant $\alpha$ on the physical properties of the strange stars were analysed.  \\

A notable body of work has emerged in the area of analysing the properties of ultracompact objects in general relativity. Horizonless spacetimes  with closed light rings (null circular geodesics) have been proposed  as possible spatially regular exotic alternatives to canonical black-hole spacetimes. Comprehensive treatments may be found in \cite{cunha,cardoso,hod2,hod3,hod4,hod5}.   In the work of   Novotný, Hladík, and Stuchlík \cite{novotny}  spherically symmetric self-gravitating isotropic ultra-compact objects were  studied numerically. It was demonstrated in \cite{hod1} that the horizonless curved spacetimes of  spatially regular compact matter  generally possess two light rings satisfying the relationship that the mass-radius ratio of the inner light ring being bounded above by the mass-radius ratio of the outer ring. Hod \cite{hod1} succeeded in generating an analytical proof for the phenomenon. The calculations were possible in the context of asymptotically flat horizonless spacetimes.In particular   self-gravitating ultra-compact trapping polytropic spheres were first considered in \cite{stuchlik}. It would be interesting to investigate the contribution of the higher curvature Gauss-Bonnet terms with regards to the bounds established in Einstein gravity.   

Realistic stellar models are expected to satisfy some elementary physical requirements \cite{knutsen}. These conditions on 4D EGB are extrapolated from the
Einstein version which are as follows. The energy
density ($\rho$) and pressure ($P$) profiles of a compact object are positive definite. The  fluid's pressure  and energy density are monotonically  decreasing outward from the centre  and the boundary defined by $P(R) = 0$ should have a real valued solution. This is used in matching of the interior metric with the   exterior  vacuum solution.   The energy conditions should also be satisfied. Here, we focused on null, weak, dominant and strong energy  conditions that can be identified as $\rho -P> 0$, $\rho +P> 0$ and $\rho +3P> 0$. Another important feature worth mentioning, is that velocity of sound should obey $0<\frac{dP}{d\rho}< 1$ to be subluminal. Chandrasekhar also established a lower bound for a certain quantity in order to guarantee the adiabatic stability of the model. All of these physical constraints will be examined once an exact model is developed.

 The plan of this paper is as follows: After the introduction in Sec. \ref{intro:Sec}, we briefly review the field equations in the 4$D$ EGB gravity and show that it makes a nontrivial contribution to gravitational dynamics in 4$D$ in Sec. \ref{sec2}.  In Sec. \ref{sec3} we note some known solutions namely the exterior metric which will be used in the matching procedure as well as the interior Schwarzschild spacetime.  We choose  particular forms for one of the gravitational potentials 
which enables us to obtain the condition of pressure isotropy in the remaining gravitational potential in Sec. \ref{sec4} and analyze the physical  properties of the model such as their energy density, pressure,  speed of sound, energy conditions and adiabatic stability. In Sec. \ref{sec5},  we give some other examples of the gravitational potentials that yield exact solutions.
Finally, in Sec. \ref{sec6}, we end with a discussion of our results and a conclusion.

 \section{Basic equations of EGB gravity} \label{sec2}
 
The Einstein-Gauss-Bonnet gravity in the $D$-dimensions spacetime can be described by the following action
\begin{equation}\label{action}
	\mathcal{I}_{G}=\frac{1}{16 \pi }\int d^{D}x\sqrt{-g}\left[ R +\frac{\alpha}{D-4} \mathcal{L}_{\text{GB}} \right]
+\mathcal{S}_{\text{matter}},
\end{equation}
where $g$ denotes the determinant of the metric $g_{\mu\nu}$ and $\alpha$ is the Gauss-Bonnet (GB) coupling constant. 
The Ricci scalar $R$  provides the GR part of the action and $\mathcal{L}_{\text{GB}}$ is the Gauss-Bonnet
Lagrangian defined as
\begin{equation}
\mathcal{L}_{\text{GB}}=R^{\mu\nu\rho\sigma} R_{\mu\nu\rho\sigma}- 4 R^{\mu\nu}R_{\mu\nu}+ R^2\label{GB}.
\end{equation}

The $\mathcal{S}_{\text{matter}}$ is the action of the matter fields. The equation of motion 
can be obtained by varying the action with respect to $g_{\mu \nu}$, as
\begin{equation}\label{GBeq}
G_{\mu\nu}+\frac{\alpha}{D-4} H_{\mu\nu}= 8 \pi  T_{\mu\nu}~,~~~\mbox{where}~~~~~~T_{\mu\nu}= -\frac{2}{\sqrt{-g}}\frac{\delta\left(\sqrt{-g}\mathcal{S}_m\right)}{\delta g^{\mu\nu}},
\end{equation}
Since, $\mathcal{L}_m$ represents the Lagrangian density of matter that depends only on the metric tensor components $g_{\mu\nu}$, and not on its derivatives.  $G_{\mu\nu}$ is the Einstein tensor with $H_{\mu\nu}$ is the contribution of the GB term with the following expression
\begin{eqnarray}
&& G_{\mu\nu} = R_{\mu\nu}-\frac{1}{2}R~ g_{\mu\nu},\nonumber\\
&& H_{\mu\nu} = 2\Bigr( R R_{\mu\nu}-2R_{\mu\sigma} {R}{^\sigma}_{\nu} -2 R_{\mu\sigma\nu\rho}{R}^{\sigma\rho} - R_{\mu\sigma\rho\delta}{R}^{\sigma\rho\delta}{_\nu}\Bigl)- \frac{1}{2}~g_{\mu\nu}~\mathcal{L}_{\text{GB}},\label{FieldEq}
\end{eqnarray}
In the above expression, $R$ is the Ricci scalar, $R_{\mu\nu}$ the Ricci tensor, $H_{\mu\nu}$  the Lanczos tensor and $R_{\mu\sigma\nu\rho}$ is the Riemann tensor, respectively. In general, $H_{\mu\nu}$ vanishes in 4$D$ space-time, and hence  does not contribute to the field equations. However, by taking the limit $D \to 4$ in (\ref{GBeq}) and by
rescaling the Gauss-Bonnet coupling constant by $ \alpha/(D-4)$, one can prevent the vanishing of $H_{\mu\nu}$ in 4$D$ space-time \cite{Glavan:2019inb}.

Here, we consider the energy-momentum tensor $T_{\mu\nu}$ as a perfect fluid source, that is
$T_{\mu\nu}=(\rho+p)u_{\mu}u_{\nu}+ pg_{\mu\nu}$, where $\rho=\rho(r)$ is the energy density of matter, 
 $p=p(r)$ is the pressure, and $u_{\nu}$ is the fluid's $D$-velocity. Now, we start by writing down the static, spherically symmetric $D$-dimensional metric  ansatz \cite{Mehdizadeh:2015jra} in the usual form:
\begin{eqnarray}\label{metric}
ds^2_{D}= - e^{2\Phi(r)}c^{2}dt^2 + e^{2\lambda(r)}dr^2 + r^{2}d\Omega_{D-2}^2,  
\end{eqnarray} 
where $d\Omega_{D-2}^2$ is the metric on the unit $(D-2)$-dimensional sphere.  As the metric is time independent and spherically symmetric, the $\Phi(r)$ and $\lambda(r)$ are
functions of the radial coordinate $r$ only. 
%%%%%%%%%%%%%%%%%%%%%%%%%%%%%%%%%%%%%%%%%%%%%%
Now, we are going to obtain the Einstein field equations in the limit $D \to 4$  for 4$D$ EGB gravity,  the $tt$, $rr$ and  $\theta \theta$ components are given by the following differential equations:
\begin{eqnarray}\label{DRE1}
&& \frac{\alpha(1-e^{-2\lambda})}{r^3}\left[4\lambda ' e^{-2\lambda}-\frac{(1-e^{-2\lambda})}{r}\right]
+e^{-2\lambda}\left({2\lambda ' \over r}-{1\over r^2}\right)+{1\over r^2}= 8 \pi \rho(r),\\
&& \frac{\alpha(1-e^{-2\lambda})}{r^3}\left[4\Phi ' e^{-2\lambda}+\frac{(1-e^{-2\lambda})}{r}\right]
+e^{-2\lambda}\left({2\Phi ' \over r}+{1\over r^2}\right)-{1\over r^2}=8 \pi p(r),\label{DRE2} \\
&&  e^{-2\lambda}\left[ \Phi ''+\Phi '^2+\frac{1}{r}\left(\Phi '-\lambda' \right) +\Phi ' \lambda' \left(\frac{8\alpha e^{-2\lambda} }{r^2}-1\right)- \frac{2\alpha (1-e^{-2\lambda})}{r^2} \right. \nonumber \\
  && \left. \times
\Bigg\{ \frac{1}{r}\left(\Phi '-\lambda' \right)-2\left(\Phi ''+\Phi '^2 -\Phi ' \lambda' \right)+ \frac{1}{r^2}(e^{2\lambda}-1) \Bigg \} \right]= 8 \pi p(r).\label{DRE3}
\end{eqnarray} 
It must be emphasized that we use the \textit{regularization} process (see Ref.  \cite{Glavan:2019inb,Cognola:2013fva}),
that leads to exactly the same black hole solutions \cite{Lu:2020iav,Hennigar:2020lsl,Casalino:2020kbt,Ma:2020ufk} at least for the case of 4$D$ spherically symmetric spacetimes. Now, we can deal with the stellar structure equations under the standard of dynamical systems. Now, combining Eq. (\ref{DRE2}) and (\ref{DRE3}), one can get the equation of pressure isotropy, which yield
\begin{eqnarray}
2\alpha\left( 2e^{-2\lambda} r^2\Phi'\lambda' +(1-e^{2\lambda}) +(1-e^{-2\lambda})(r^2 W + r(\lambda' -3\Phi') +1) \right)\nonumber \\ 
+r^2\left(r^2W -r(\phi' + \lambda') +(e^{2\lambda}-1)   \right)  =0, \label{iso}
\end{eqnarray}
where we have defined $W = \Phi'' + \Phi^2 -\Phi' \lambda'$. Of course, the conservation equation of the energy-momentum tensor,
$T^{\mu\nu}_{;\mu} = 0$ that for a perfect fluid is
\begin{eqnarray}\label{tov}
\frac{dp}{dr}= -(\rho + p) \frac{d\Phi}{dr} ,
\end{eqnarray}

At this stage we introduce a coordinate  transformation 
\begin{eqnarray}
x = Cr^2,~~ e^{-2\lambda} = Z(x)~~~ \mathrm{and}~~~ e^{2\Phi} = y^2(x), \label{ts}
\end{eqnarray}
where  $C$ is an arbitrary constant. The benefit of this transformation is that the master nonlinear  isotropy field equation is transformed to a linear differential equation and we can profit from the vast knowledge on such equations.  With this choice of (\ref{ts}),  we are able to express the components of the
field Eqs. (\ref{DRE1}), (\ref{DRE2}) and (\ref{iso}) can be written as
\begin{eqnarray}
8\pi \rho &=& \frac{C}{2 x^2}  \bigg[4 x (\beta (Z-1) -x) \dot{Z} - (Z-1) (2 x+\beta  (Z-1))\bigg], \label{100a}\\
8\pi p &=& \frac{C}{2 x^2 y} \bigg[ 8 x Z \dot{y} (x-\beta ( Z-1))+y (Z-1) (2 x+\beta  (Z-1)) \bigg],\\
0&=& \beta \left(4x^2Z(1-Z)\ddot{y} + 2x\left[2Z(Z - 1) + x\dot{Z}(1-3Z)\right] \dot{y}  -(1-Z)(\dot{Z}x -Z +1)y \right) + 
\nonumber \\ &&
x(4x^2 Z\ddot{y} + 2x^2\dot{Z}\dot{y} + (\dot{Z}x -Z + 1)y) , \label{100} 
\end{eqnarray}
where dots denote differentiation with respect to the variable $x$, 
which is  quasilinear. Here we define $\beta = 2 \alpha C$.  It is a second order linear differential equation in the variable $y$ and first order Abel equation of the second kind when written in terms of $Z$.  The field equation bears a striking resemblance to the structure of the 6 dimensional Einstein--Gauss--Bonnet equation of pressure isotropy but not the five dimensional version.  When the Gauss-Bonnet coupling $\beta$ vanishes the four dimensional equation of pressure isotropy in Einstein gravity 
\begin{eqnarray}
4 x^2 Z \ddot{y} + 2 x^2  \dot{Z} \dot{y} +  \left(x \dot{Z}-Z+1\right) y = 0,
\end{eqnarray}
is regained. 
It is also interesting to draw a comparison with the standard EGB equations. For five dimensional EGB gravity using the metric (\ref{metric}) the pressure isotropy equation has the form 
\beq
 0 &=& 2\beta \left( 2xZ(1-Z)\ddot{y} + (2Z(Z-1) + x\dot{Z} (1-3Z))\dot{y}\right)  + \left(2x^2Z\ddot{y} +x^2 \dot{Z} \dot{y} + (\dot{Z}x -Z +1)y \right), \label{10d}
\eeq
for an unscaled coupling parameter $\alpha$ while in six dimensional EGB gravity the master isotropy equation is given by 
\beq
0 &=& 6\beta \left(4x^2Z(1-Z)\ddot{y} + 2x(2Z(Z-1) + \dot{Z}x (1-3Z))\dot{y} + (1-Z)(\dot{Z}x -Z + 1)y  \right) \nonumber \\ && + x\left( 4x^2 Z\ddot{y} + 2x^2 \dot{Z}\dot{y} + 3(\dot{Z}x -Z+1)y \right) ,  \label{10e}
\eeq
after inserting a suitable extra angular dependence in (\ref{metric}).

\section{Known solutions}\label{sec3}

In order to establish the applicable exterior metric we may set $\rho = 0$ in (\ref{100a}). The solution of the differential equation is then given by
\beq
Z =  1 +\frac{x}{\beta} \left( 1 \pm \sqrt{1 + \frac{\beta C_1}{x^{\frac{3}{2}}}}\right), \label{100h}
\eeq
where $C_1$ is an integration constant. This corroborates the result of Doneva and Yazadjiev \cite{Doneva:2020ped}, (see \cite{Ghosh:2020syx} for black hole solution). Note that in the limit of $\beta $ approaching zero, the Schwarzschild exterior solution is regained from (\ref{100h}) but only for the negative branch of the solution. The limit otherwise is indeterminate. This can be seen by a series expansion of the square root term in powers of $\beta$. Following Boulware and Deser \cite{Boulware:1986dr} we may write (\ref{100h}) in the form 
\be
e^{-2\lambda} = 1 + \frac{r^2}{2\alpha} \left(1 - \sqrt{1 + \frac{8M \alpha}{r^3}}   \right), \label{100c}
\ee
which will be useful during the matching of the interior and exterior spacetime in order to settle the values of the constant $C$ and any other integration constant.   Note that the coefficient $\frac{ C_1}{C^{{\frac{3}{2}}}}$ is identified with the mass $M$ of the star. Observe also that the solution (\ref{100h}) may also be obtained by setting $y = \sqrt{Z}$ in isotropy equation (\ref{100}) since it is well known that vacuum solutions have the form 
\be
ds^2 = - F(r) dt^2 + \frac{1}{F(r)} dr^2 + r^2 d\Omega^2 ,
\ee 
for some function $F(r)$ and this serves as a useful check on the veracity of the field equations. 

Another known solution is that of the  constant density or incompressible fluid sphere. This was studied in  \cite{Doneva:2020ped}, however the solution obtained is precisely the form of the Schwarschild interior metric of classical GR. This supports the conclusion of Dadhich {\it{et al}} who proved that the Schwarzschild interior solution is universal in EGB in all dimensions and evidently this is true for the 4$D$ scenario. We can easily verify this by putting $Z = 1 + x$ into field equation (\ref{100}) and obtaining the potential
\be
 y = A + B\sqrt{1+x}, \label{100d}
\ee
where $A$ and $B$ are constants of integration. Essentially for the constant density value $\rho = \rho_0$ the spatial potential may be written as
\be
e^{-2\lambda} = 1 + \frac{r^2}{2\alpha} \left(1 - \sqrt{1 + \frac{8\rho_0 \alpha}{3}}   \right)  ,
\ee
as reported in \cite{Doneva:2020ped,Banerjee:2020stc} and which is clearly of Schwarzschild form $Z = e^{-2\lambda} = 1 + Cr^2$.

%%%%%%%%%%%%%%%%%%%%%%%%%%%%%%%%%%%%%%%%%%%%%
\section{A physically reasonable exact model}\label{sec4}

Generating exact solutions to the equations of motion comprising the Einstein equations supplemented by the Gauss--Bonnet contributions is a nontrivial matter. Historically, the Einstein equations in the case of isotropic perfect fluids, have admitted some 120 exact solutions most of which were found through mathematical motivations. This is because the system of field equations is underdetermined and requires a further stipulation to close the system. Implementing physically relevant conditions such as specifying an equation of state such as a linear barotropic or polytropic one has drawn a blank as respects exact solutions. Each of these cases were solved numerically by Nillson and Uggla \cite{nilsson1,nilsson2}. In general, the freedom of choice has often been exercised with the selection of a suitable potential ansatz and by integration determining the remaining potential and all remaining physical quantities. {\it{Post-facto}} it is then checked to see whether an equation of state $p = p(\rho)$ could be explicitly determined. A notable exception to this approach was that of Tolman \cite{tolman}. By a cunning rearrangement of the master isotropy equation, he was able to determine eight classes of solutions by considering the vanishing of certain parts of the equation. Five of these turned out to be new and some such as Tolman IV and Tolman VII have been shown to be physically viable. Implementing the same scheme in our case does not yield palatable results. In fact inserting the Tolman IV and VII prescriptions into the isotropy equation (\ref{90a}) leads to an intractable differential equation. The approach we follow has been successful in finding exact solutions to the five dimensional EGB system and constitute the only ones detectable to the present time.

The equation (\ref{100}) may be converted to the form 
\begin{eqnarray} 
\beta \left( x(2x (1-3Z)\dot{y} + y (Z-1))\dot{Z} - (4x^2\ddot{y} - 4x^2 \dot{y} +y)Z^2 + (4x^2 \ddot{y} -4x\dot{y} +2y)Z -y \right) \nonumber && \\        + x \left(x(2x\dot{y} +y)\dot{Z} +(4x^2 \ddot{y} -y)Z + y   \right) &=& 0,    \label{90a}
\end{eqnarray}
by rearrangement. Viewed as a differential equation in $Z(x)$ equation (\ref{90a}) is an Abel equation of the second kind with very few classes of known solutions expressible as elementary functions.  In 
order to determine an exact solution we prescribe the temporal potential to have the form $y = \sqrt{x}$. This choice is made on mathematical grounds and corresponds to a case of the Tolman VI ansatz. This approach has been followed repeatedly historically  in finding exact solutions of the much simpler Einstein equations as opposed to the method of imposing a physical prescription such as an equation of state. This latter approach generally ends in resorting to the use of numerical techniques with their accompanying approximations. Our objective is to detect an exact solution and thereafter to deduce the physical characteristics of the model. The one drawback of this choice of metric potential is that at the stellar centre $x = 0$ it is not defined. This suggests that if all other physical requirements are met, then our model may model a shell of perfect fluid surrounding a core with well behaved physical properties at the centre. This phenomenon is bound to show up in what comes next in the form of singularities in the physical quantities which are not removable. 

Inserting $y = \sqrt{x}$ in (\ref{90a}) the  real valued function
\begin{eqnarray} 
Z = \frac{1}{2}\left(\frac{\sqrt[3]{-2 k{}^6 x^3+2 \sqrt{k{}^{15} x^3+k{}^{12} x^6}-k{}^9}}{k{}^3}+\frac{k{}^3}{\sqrt[3]{-2 k{}^6 x^3+2 \sqrt{k{}^{15} x^3+k{}^{12} x^6}-k{}^9}}+2 x+1\right),
\end{eqnarray} 
where $k$ is a constant of integration and two complex valued potentials for $\beta = 1$ emerge.The complex roots are discarded as they have no physical meaning in the context of stellar structure.  At this stage of computational capability the general solution for all $\beta$ is not available, but it is pleasing that we have a nontrivial value of the coupling constant present so that we can study the influence of the higher curvature terms in the gravitational behaviour of the four dimensional spherically symmetric spacetime.  

The density $\rho$ and isotropic particle pressure $p$ evaluate to 
\begin{eqnarray} 
8\pi\rho &=& \frac{ C}{2 x^2}\left(x \left(k^3 x^2 \left(k^6-v_2^{2/3}\right)+2 v_1 \sqrt[3]{v_2}\right)\left(\beta  k^6+k^3 \sqrt[3]{v_2} \left(2 v_3 x-\beta \right)+\beta  v_2^{2/3}\right)\right. / \nonumber\\
&&\left.k^3 v_1 v_2^{2/3}-\frac{1}{2} \left(\frac{\sqrt[3]{v_2}}{k^3}+\frac{k^3}{\sqrt[3]{v_2}}+2 x-1\right) \left(\frac{1}{2} \beta  \left(\frac{\sqrt[3]{v_2}}{k^3}+\frac{k^3}{\sqrt[3]{v_2}}-1\right)+(\beta +2) x\right)\right), \\
8\pi p &=& \frac{ C}{2 x^2} \left(\frac{1}{4} \beta  \left(-\frac{3 v_2^{2/3}}{k^6}-\frac{3 k^6}{v_2^{2/3}}-\frac{2 \sqrt[3]{v_2}}{k^3}-\frac{2 k^3}{\sqrt[3]{v_2}}-1\right)-\frac{v_3 x \left(3 k^6+k^3 \sqrt[3]{v_2}+3 v_2^{2/3}\right)}{k^3 \sqrt[3]{v_2}}-3 (\beta -2) x^2\right),
\end{eqnarray} 
where we have  applied the  substitutions 
$v_1 = \sqrt{k^{12} x^3 \left(k^3+x^3\right)}$ , $v_2 = -k^9-2 k^6 x^3+2 v_1$ and $v_3 = \beta -1$.In light of the fact that both density and pressure have very complicated forms, the determining of a barotropic equation of state $p = p(\rho)$ becomes intractable. Consequently we will rely upon the quantity $\frac{p}{\rho}$ to convey an idea of the variation of the equation of state.  
The square of the sound speed  has the complicated  form 
\begin{eqnarray}
\frac{dp}{d\rho} &=& -\left(\left(k^3+x^3\right)\left(2 k^{15} x^3 \left(9 v_3 x-\beta \right)+2 v_1 v_2^{2/3} x^3 \left(\beta +2 v_3 x\right)+2 k^{12} x^3 \left(2 \beta  \sqrt[3]{v_2}+12 v_3 x^4-9 v_3 \sqrt[3]{v_2} x+2 \beta  x^3\right)\right.\right. \nonumber \\
&& +k^3 v_1 \left(\beta  v_2^{2/3}+24 v_3 \sqrt[3]{v_2} x^4-6 \beta  \sqrt[3]{v_2} x^3+2 v_3 v_2^{2/3} x-8 \beta  x^6\right) \nonumber\\
&& +k^9 \left(\beta  v_1-24 v_3 \sqrt[3]{v_2} x^7+6 \beta  \sqrt[3]{v_2} x^6-4 v_3 v_2^{2/3} x^4-2 \beta  v_2^{2/3} x^3-6 v_1 v_3 x+8 \beta  x^9\right) \nonumber\\
&& \left.\left.-k^6 \left(\beta  v_1 \sqrt[3]{v_2}+4 v_3 v_2^{2/3} x^7+2 \beta  v_2^{2/3} x^6+24 v_1 v_3 x^4-6 v_1 v_3 \sqrt[3]{v_2} x\right)\right)\right) / \nonumber \\
&& \left(2 k^{18} x^3 \left(5 \beta +2 v_3 x\right)-2 k^{15} x^3 \left(4 \beta  \sqrt[3]{v_2}+7 v_3 x^4+4 v_3 \sqrt[3]{v_2} x-20 \beta  x^3\right)-2 v_1 v_2^{2/3} x^6 \left(3 \beta +2 v_3 x\right)\right. \nonumber \\
&& +k^3 v_1 x^3 \left(-9 \beta  v_2^{2/3}-8 v_3 \sqrt[3]{v_2} x^4+30 \beta  \sqrt[3]{v_2} x^3-6 v_3 v_2^{2/3} x+8 \beta  x^6\right) \nonumber \\
&& +k^{12} \left(-3 \beta  v_1-24 v_3 x^{10}-6 v_3 \sqrt[3]{v_2} x^7-32 \beta  \sqrt[3]{v_2} x^6+4 v_3 v_2^{2/3} x^4+6 \beta  v_2^{2/3} x^3-2 v_1 v_3 x+28 \beta  x^9\right) \nonumber \\
&& +k^9 \left(3 \beta  v_1 \sqrt[3]{v_2}+8 v_3 \sqrt[3]{v_2} x^{10}-30 \beta  \sqrt[3]{v_2} x^9+8 v_3 v_2^{2/3} x^7+12 \beta  v_2^{2/3} x^6+2 v_1 v_3 x^4-23 \beta  v_1 x^3+2 v_1 v_3 \sqrt[3]{v_2} x-8 \beta  x^{12}\right) \nonumber \\
&& \left.+k^6 \left(-3 \beta  v_1 v_2^{2/3}+4 v_3 v_2^{2/3} x^{10}+6 \beta  v_2^{2/3} x^9+24 v_1 v_3 x^7-32 \beta  v_1 x^6+10 v_1 v_3 \sqrt[3]{v_2} x^4+17 \beta  v_1 \sqrt[3]{v_2} x^3-2 v_1 v_3 v_2^{2/3} x\right)\right), \nonumber \\  \label{sp}
\end{eqnarray}
while the expressions 
\begin{eqnarray}
\rho - p &=& \frac{ C}{32 \pi v_1 v_2^{2/3} x^2} \left(4 k^{12} x^3 \left(\beta +5 v_3 x\right)+2 k^9 x^3 \left(\beta  \sqrt[3]{v_2}+12 v_3 x^4+2 v_3 \sqrt[3]{v_2} x\right)+k^6 \left(-\beta  v_1+6 \beta  \sqrt[3]{v_2} x^6-8 v_1 v_3 x\right) \right. \nonumber \\
&& \left. +k^3 v_1 \left(\beta  \sqrt[3]{v_2}-24 v_3 x^4+8 v_3 \sqrt[3]{v_2} x\right)-v_1 \sqrt[3]{v_2} \left(\beta  \sqrt[3]{v_2}-12 (\beta -2) \sqrt[3]{v_2} x^2+6 \beta  x^3\right)\right), \\
\rho + p &=& \frac{C}{16 \pi v_1^{5/3} x^2} \left(-2 k^{12} v_1 x^4+k^6 v_1 \left(2 v_1 x-\beta \right)-v_1^{2/3} v_1 \left(\beta +2 v_1 x\right)\right. \nonumber\\
&& \left.\left.k^3 v_1 \left(\beta  \sqrt[3]{v_1}-2 \sqrt[3]{v_1} v_1 x+2 \beta  x^3\right)-2 k^9 x^3 \left(\beta  \sqrt[3]{v_1}-\sqrt[3]{v_1} v_1 x+\beta  x^3\right)\right)\right),\\
\rho + 3p &=& \frac{ C}{32 \pi v_1^{5/3} x^2} \left(v_1 \sqrt[3]{v_1} \left(-3 \beta  \sqrt[3]{v_1}-12 (\beta -2) \sqrt[3]{v_1} x^2-8 v_1 \sqrt[3]{v_1} x+6 \beta  x^3\right)-4 k^{12} x^3 \left(\beta +7 v_1 x\right)\right. \nonumber\\
&& -2 k^9 x^3 \left(5 \beta  \sqrt[3]{v_1}+12 v_1 x^4-2 v_1 \sqrt[3]{v_1} x+4 \beta  x^3\right)+k^6 \left(-3 \beta  v_1-6 \beta  \sqrt[3]{v_1} x^6+16 v_1 v_1 x\right) \nonumber \\ 
&&\left.\left.+k^3 v_1 \left(3 \beta  \sqrt[3]{v_1}+24 v_1 x^4-16 v_1 \sqrt[3]{v_1} x+8 \beta  x^3\right)\right)\right),
\end{eqnarray}
will be useful in studying the energy conditions. 
Finally we exhibit the Chandrasekhar adiabatic stability index
\begin{eqnarray} 
\Gamma &=& -\left(\left(4v_1\left(2 k^{12} v_3 x^4+2 k^9 x^3 \left(\beta  \sqrt[3]{v_2}-v_3 \sqrt[3]{v_2} x+\beta  x^3\right)-k^6 v_1 \left(2 v_3 x-\beta \right)+v_1 v_2^{2/3} \left(\beta +2 v_3 x\right)\right.\right.\right. \nonumber\\
&& \left.-k^3 v_1 \left(\beta  \sqrt[3]{v_2}-2 v_3 \sqrt[3]{v_2} x+2 \beta  x^3\right)\right)\left(2 k^{15} x^3 \left(9 v_3 x-\beta \right)+2 v_1 v_2^{2/3} x^3 \left(\beta +2 v_3 x\right)\right. \nonumber\\
&& +2 k^{12} x^3 \left(2 \beta  \sqrt[3]{v_2}+12 v_3 x^4-9 v_3 \sqrt[3]{v_2} x+2 \beta  x^3\right) \nonumber\\
&& -k^3 v_1 \left(-\beta  v_2^{2/3}-24 v_3 \sqrt[3]{v_2} x^4+6 \beta  \sqrt[3]{v_2} x^3-2 v_3 v_2^{2/3} x+8 \beta  x^6\right) \nonumber\\
&& -k^9 \left(-\beta  v_1+24 v_3 \sqrt[3]{v_2} x^7-6 \beta  \sqrt[3]{v_2} x^6+4 v_3 v_2^{2/3} x^4+2 \beta  v_2^{2/3} x^3+6 v_1 v_3 x-8 \beta  x^9\right)  \nonumber\\
&& \left.\left. -k^6 \left(\beta  v_1 \sqrt[3]{v_2}+4 v_3 v_2^{2/3} x^7+2 \beta  v_2^{2/3} x^6+24 v_1 v_3 x^4-6 v_1 v_3 \sqrt[3]{v_2} x\right)\right)\right) / \nonumber\\
&& \left(k^6x^3\left(6 \beta  v_1 \sqrt[3]{v_2}+k^{12} \left(\beta -12 v_3 x\right)+4 k^3 v_1 \left(\beta +6 v_3 x\right)+k^9 \left(-\beta  \sqrt[3]{v_2}-24 v_3 x^4+12 v_3 \sqrt[3]{v_2} x-4 \beta  x^3\right)\right.\right. \nonumber\\
&& \left.+k^6 \left(\beta  v_2^{2/3}-6 \beta  \sqrt[3]{v_2} x^3+12 (\beta -2) v_2^{2/3} x^2+4 v_3 v_2^{2/3} x\right)\right) \nonumber\\
&& \times \left(2 k^{18} x^3 \left(5 \beta +2 v_3 x\right)-2 k^{15} x^3 \left(4 \beta  \sqrt[3]{v_2}+7 v_3 x^4+4 v_3 \sqrt[3]{v_2} x-20 \beta  x^3\right)-2 v_1 v_2^{2/3} x^6 \left(3 \beta +2 v_3 x\right)\right. \nonumber\\
&&+k^3 v_1 x^3 \left(-9 \beta  v_2^{2/3}-8 v_3 \sqrt[3]{v_2} x^4+30 \beta  \sqrt[3]{v_2} x^3-6 v_3 v_2^{2/3} x+8 \beta  x^6\right) \nonumber\\
&& -k^{12} \left(3 \beta  v_1+24 v_3 x^{10}+6 v_3 \sqrt[3]{v_2} x^7+32 \beta  \sqrt[3]{v_2} x^6-4 v_3 v_2^{2/3} x^4-6 \beta  v_2^{2/3} x^3+2 v_1 v_3 x-28 \beta  x^9\right)  \nonumber\\
&& +k^9 \left(3 \beta  v_1 \sqrt[3]{v_2}+8 v_3 \sqrt[3]{v_2} x^{10}-30 \beta  \sqrt[3]{v_2} x^9+8 v_3 v_2^{2/3} x^7+12 \beta  v_2^{2/3} x^6 \right)  \nonumber\\
&&+k^9 \left( 2 v_1 v_3 x^4-23 \beta  v_1 x^3+2 v_1 v_3 \sqrt[3]{v_2} x-8 \beta  x^{12}\right) \nonumber\\
&& +k^6\left(4 v_3 v_2^{2/3} x^{10}+24 v_1 v_3 x^7+10 v_1 v_3 \sqrt[3]{v_2} x^4-2 v_1 v_3 v_2^{2/3} x\right) \nonumber\\ 
&& \left.\left.\left.+k^6\left(-3 \beta  v_1 v_2^{2/3}+6 \beta  v_2^{2/3} x^9-32 \beta  v_1 x^6+17 \beta  v_1 \sqrt[3]{v_2} x^3\right)\right)\right)\right),
\end{eqnarray} 
is established with the help of the formula $\Gamma = \left(\frac{\rho+p}{p}\right)\frac{dp}{d\rho}$. These lengthy and complicated expressions are common in the study of higher curvature effects induced by the Gauss--Bonnet contribution. Clearly an analytic investigation of these physical quantities lies outside the realm of possibility and so we endeavour to analyse the effect of the higher curvature terms through graphical means.

\subsection{Qualitative physical analysis }\label{sec5}

We now consider the dynamical quantities graphically  using the parameter values $k = -1$, $\beta = - 1$ and $C = 1$. These parameter values have been so selected through a careful and tedious process of empirical fine tuning until physically reasonable behaviour of the dynamical variables emerged. Observe there is no reason to demand a positive coupling constant $\beta = 2\alpha C$. It is known that $\alpha$ is connected to the string tension in string theory and so is expected to be positive there, however, it has a different meaning as a coupling parameter in the present gravitational theory.  Additionally, for the same parameter values, we have plotted the Einstein only ($\beta = 0$) on the same system of axes to facilitate a comparison and to examine the influence of the GB terms. 

  \begin{itemize}
  \item{Density and pressure profiles}
  
  At the very least it is demanded that both density and pressure be positive definite functions that are monotonically decreasing outwards from the stellar centre.  This is clearly in evidence in both figures 1 and 2. Moreover from figure 2, it is clear that the pressure vanishes for some finite radius at roughly $x = 0.085$ in the GB case and at a higher value in the Einstein case. This suggests that the higher curvature terms have the effect of decreasing the radius of the sphere. The density plot in figure 1 also indicates that the Einstein counterpart is not physically viable being negative everywhere. The additional GB terms appear to correct this undesirable behaviour. 

\begin{figure}[h]
	\centering
	\begin{subfigure}{0.7\textwidth}
		\includegraphics[width=0.9\linewidth]{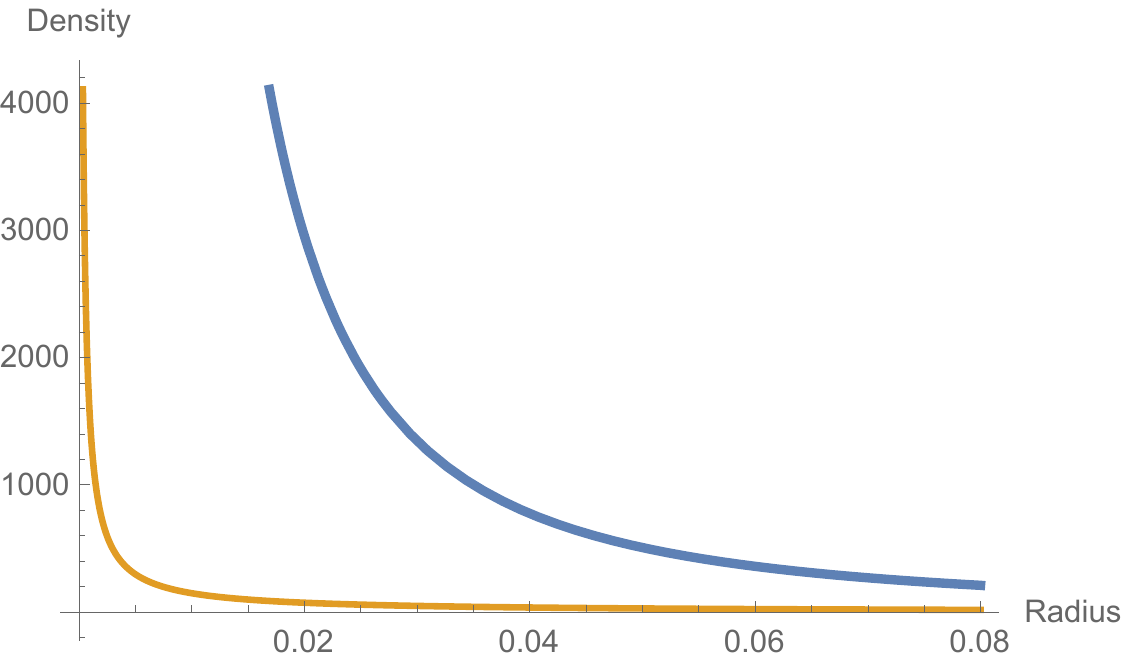}
		\label{fig:sub1}
	\end{subfigure}\hspace*{-3.5em}%
	\begin{subfigure}{0.35\textwidth}
		\includegraphics[width=0.4\linewidth]{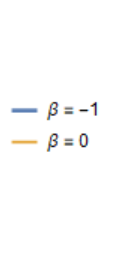}
		\label{fig:sub2}
	\end{subfigure}
	\caption[Density]{Variation of density $\rho$ against the radial parameter $x$. }
	\label{fig:density}
\end{figure}

\begin{figure}[h]
	\centering
	\begin{subfigure}{0.7\textwidth}
		\includegraphics[width=0.9\linewidth]{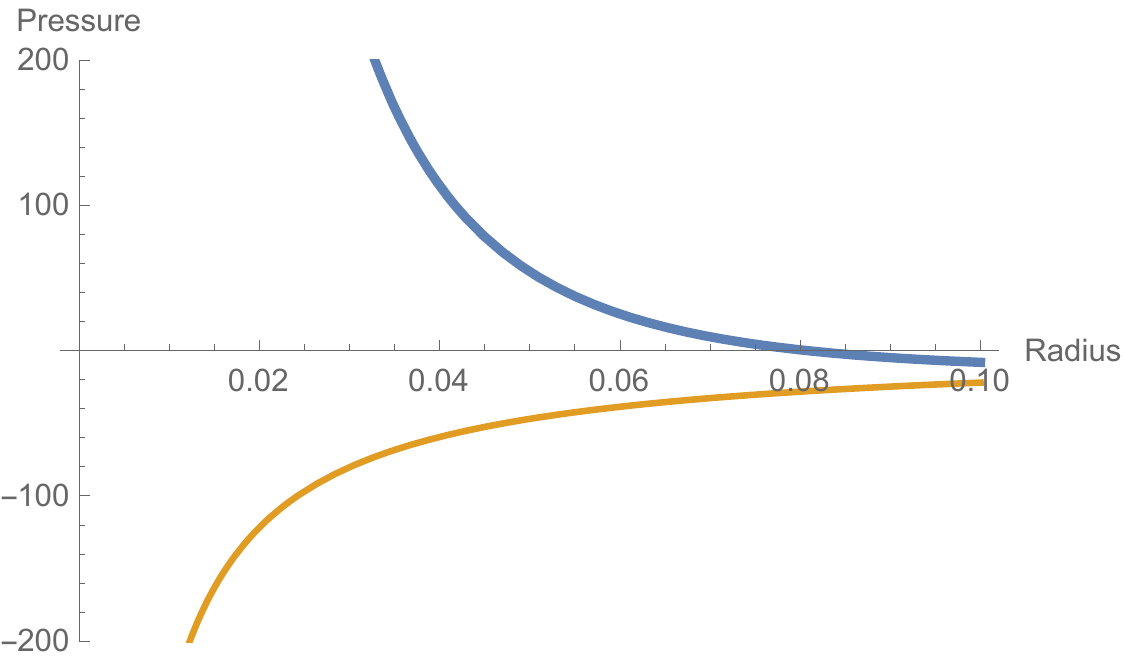}
		\label{fig:sub1}
	\end{subfigure}\hspace*{-3.5em}%
	\begin{subfigure}{0.35\textwidth}
		\includegraphics[width=0.4\linewidth]{key2.png}
		\label{fig:sub2}
	\end{subfigure}
	\caption[Pressure]{Variation of pressure $p$ against the radial parameter $x$. }
	\label{fig:pressure}
\end{figure}

\item{Energy conditions}

In order to investigate further the structure and property of our model, we focus on the energy conditions (ECs). It is reasonably expected   that every type of classical matter
satisfies the energy conditions whether in modified gravity or standard general relativity. There are several  energy conditions
that are  sets of inequalities depending on the energy-momentum tensor of matter. The most fundamental of these is  the weak energy condition (WEC), i.e. $T_{\mu\nu} U^{\mu}U^{\nu}$ $\geq 0$, where $U^{\mu}$ is a timelike vector. For the given diagonal energy momentum tensor, the WEC requires that,
\begin{equation}\label{EC1}
\rho (r)\geq 0 ~~\text{and}~~ \rho (r)+ p(r)\geq 0.
\end{equation}
Indeed, continuity of the WEC implies the null energy condition which requires 
 for any null vector $k^{\mu}$ that  $T_{\mu\nu} k^{\mu} k^{\nu}\geq 0$. The strong energy condition (SEC) is given by,
 $\left(T_{\mu\nu}-\frac{1}{2} T g_{\mu\nu}\right) U^{\mu} U^{\nu}\geq 0$ for any  timelike vector $U^{\mu}$, where $T$ is the trace of the stress energy tensor. In terms of the diagonal stress energy tensor the SEC reads
\begin{equation}\label{EC2}
\rho (r)+ \sum p_{i}(r)\geq 0 ~~\implies~~ \rho (r)+3 p(r)\geq 0.
\end{equation}
Finally the dominant energy condition (DEC)  demands  that for any timelike vector $U^{\mu}$ that $T_{\mu\nu} U^{\mu}U^{\nu}$ $\geq 0$ and $T_{\mu\nu} U^{\nu}$ is not spacelike. Accordingly the constraints 
\begin{equation}\label{EC3}
\rho (r)\geq 0 ~~\text{and}~~ \rho (r)- |p(r)| \geq 0.
\end{equation}
arise. The results obtained from our requirements (\ref{EC1} -- \ref{EC3})  are illustrated
in Fig. \ref{fig:energyconditions}, for $\beta= -1$ and $\beta= 0$. These plots imply that all energy conditions are satisfied in the 4D EGB case however  the DEC is violated in the Einstein model when $\beta=0$. The null energy condition is verified in Fig 1 for both standard and modified models. \ref{fig:density}.

\begin{figure}[h]
	\centering
	\begin{subfigure}{0.7\textwidth}
		\includegraphics[width=0.9\linewidth]{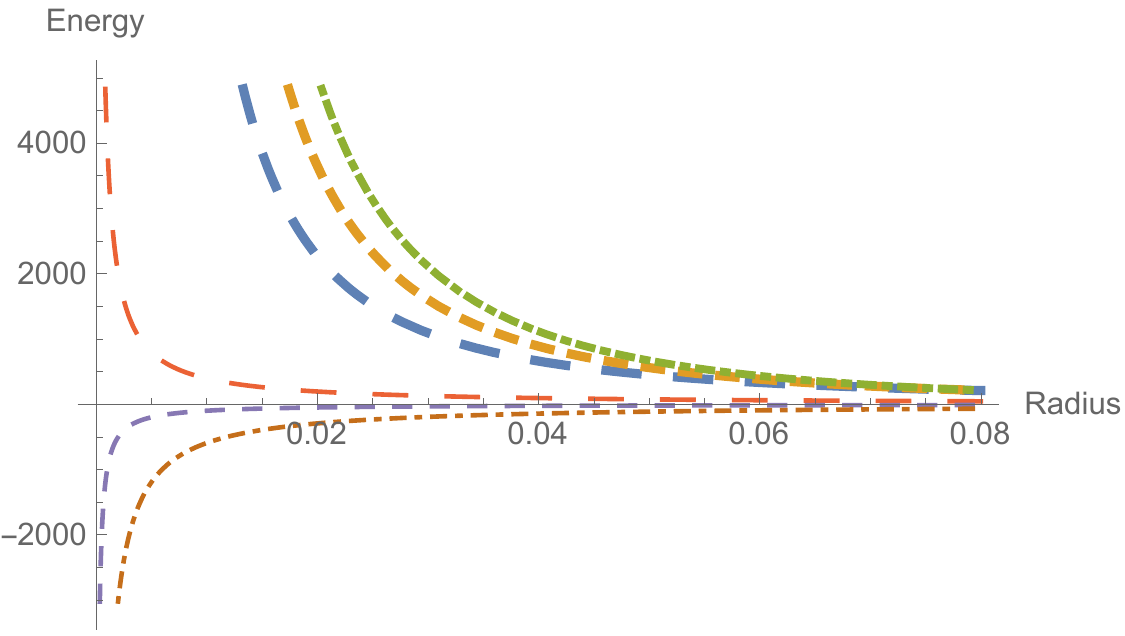}
		\label{fig:sub1}
	\end{subfigure}\hspace*{-3.8em}%
	\begin{subfigure}{0.55\textwidth}
		\includegraphics[width=0.4\linewidth]{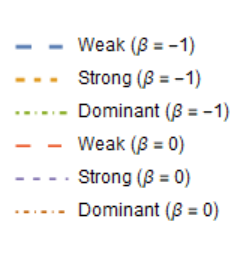}
		\label{fig:sub2}
	\end{subfigure}
	\caption[Energy Conditions]{Variation of  energy conditions versus the radial variable $x$.}
	\label{fig:energyconditions}
\end{figure}

\item{Sound speed and  causality}

 In order for the causality to be preserved, it is natural to require that the sound speed   does not exceed the speed of light, i.e.  $0 \leq v^2_s = dp/ d\rho <1$ (in geometrical units). Here, we  investigate the speed of sound propagation using the expression \eqref{sp}. From figure 3 it is observed that the sound speed is always subluminal ($0 < \frac{dp}{d\rho} < 1$) for the GB ($\beta = 1$) case within the radial value $x = 0.08$  but fails totally for the Einstein case ($\beta = 0$). We may infer that the Gauss--Bonnet higher curvature terms were responsible for the correction in behaviour of the Einstein model.

\begin{figure}[h]
	\centering
	\begin{subfigure}{0.7\textwidth}
		\includegraphics[width=0.9\linewidth]{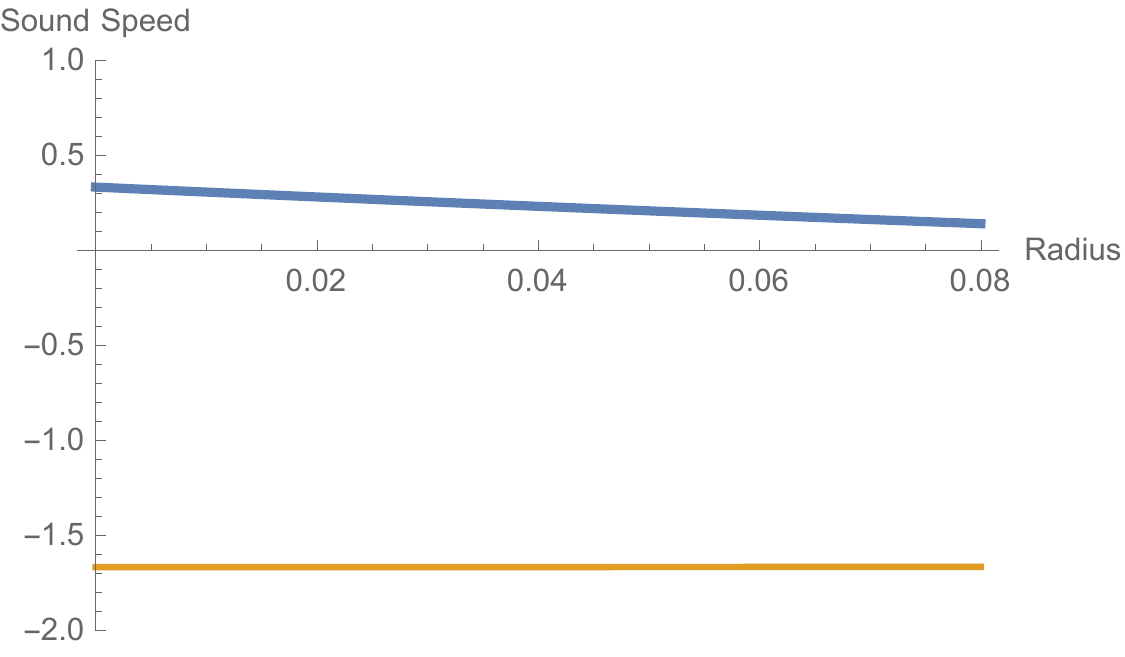}
		\label{fig:sub1}
	\end{subfigure}\hspace*{-3.5em}%
	\begin{subfigure}{0.35\textwidth}
		\includegraphics[width=0.4\linewidth]{key2.png}
		\label{fig:sub2}
	\end{subfigure}
\caption[Sound Speed Index]{Square of sound speed $\frac{dp}{d\rho}$ versus the radial variable $x$.}
	\label{fig:soundspeed}
\end{figure}

\item{Chandrasekhar adiabatic stability criterion }

The dynamical stability of a compact is computed through the  adiabatic index ($\gamma$) which is an important thermodynamical quantity. It has been discussed by Chandrasekhar \cite{Chandrasekhar} who considered the instability problem based on the variational method. The usual representations of the high-density EoS of perfect fluid 
are based on parameterizing the adiabatic index $\gamma$ defined by
\begin{equation}
\gamma \equiv \left(1+\frac{\rho}{p}\right)\left(\frac{dp}{d\rho}\right)_S,
\end{equation}
where the derivation is performed at constant entropy $S$ and $dp/ d\rho$   is the speed of sound in units of the speed of light. Thus, the sound speed is an important quantity  closely related to the adiabatic index. For instance, for the Schwarzschild star with constant-density, the  $\gamma$  for the  fluid is infinite (incompressible  fluid). However, in Ref. \cite{Glass} the authors  have shown that the adiabatic index should exceed $4/3$ for a stable polytropic star by an amount that depends on that ratio $\rho/p$ at the centre of the star. But from recent observations it is suggested  that the range of $\gamma$  should be  2 to 4 in most  neutron stars \cite{Haensel}. In Fig. \ref{fig:adiabaticindex} we plot 
$\gamma$ as a function of radial distance. These results are once again consistent  with expectations i.e.  $\gamma> 4/3$.

\begin{figure}[h!]
	\centering
	\begin{subfigure}{0.7\textwidth}
		\includegraphics[width=0.9\linewidth]{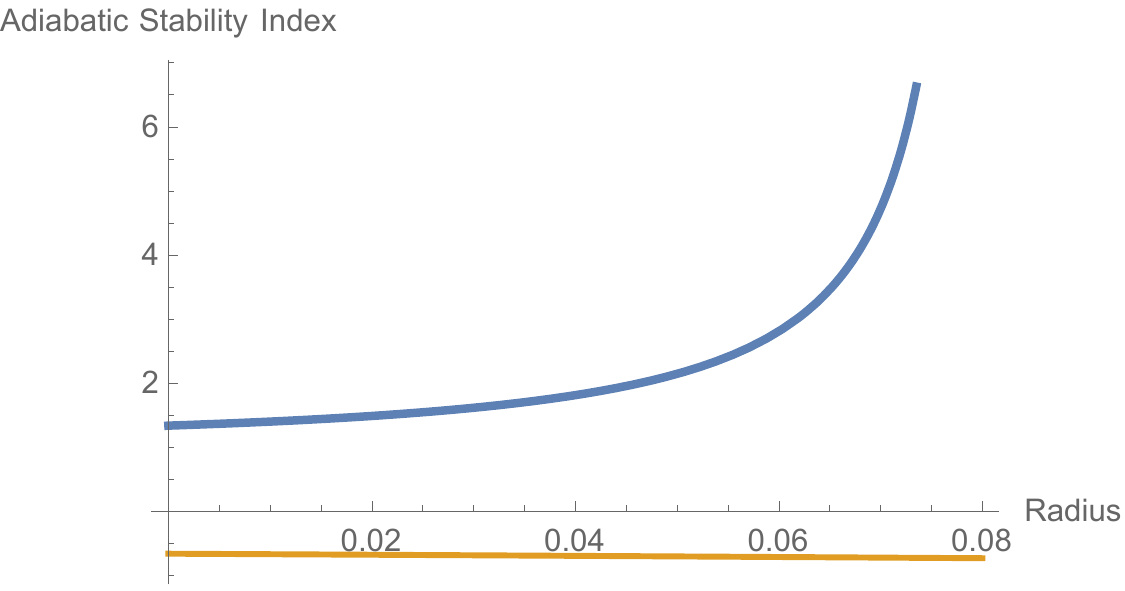}
		\label{fig:sub1}
	\end{subfigure}\hspace*{-3.5em}%
	\begin{subfigure}{0.35\textwidth}
		\includegraphics[width=0.4\linewidth]{key2.png}
		\label{fig:sub2}
	\end{subfigure}
	\caption[Chandrasekhar Adiabatic Stability Index]{Chandrasekhar adiabatic stability index versus the radial variable $x$.}
	\label{fig:adiabaticindex}
\end{figure}

\item{Matching}

The line element for the star at the boundary $(r=R)$ has the form 
\begin{eqnarray}
{ds}^2 &=& -\left(CR^{2} \right) {dt}^2 + \frac{2}{  \left(
\frac{1}{\eta}  +\eta +2 {CR}^2+1 \right)} dr^2 +r^2(d\theta^2 + \sin^2 \theta d\phi^2), \label{1000}
\end{eqnarray}
where we have set  $\eta= \frac{k^3}{\sqrt[3]{-2 \left({CR}^2\right)^3 k^6+2 \sqrt{\left({CR}^2\right)^3 k^{15}+\left({CR}^2\right)^6 k^{12}}-k^9}}$. The metric 
(\ref{1000}) should be  matched with the exterior line element (\ref{100c}). Accordingly the arbitrary constant  $C$ may be expressed as
\begin{eqnarray}
C = \frac{1}{R^2}\left( 1-\frac{2M}{R} \right) 
\end{eqnarray}
in terms of the radius $R$ and mass $M$ of the sphere. From the matching of the $g_{11}$ component 
the integration constant $k$ may be established as  

\begin{eqnarray}
k &=&  C R^2 \left(\zeta_2  +\zeta_3 \right)\bigg/  \sqrt[3]{2} \zeta_1  \left(-\left(1+\frac{1}{8 \zeta_1^3}\left(\zeta_2 +\zeta^2_3\right)^3\right)^2\right)^\frac{1}{3}, 
\end{eqnarray}
where we have set
$\zeta_1= \left(\alpha -2 M R+R^2\right),$
$\zeta_2= \alpha +4 C M R^3-2 C R^4-2 \alpha  C R^2+R^2 \sqrt{\frac{8 \alpha  M}{R^3}+1}+2 M R,$

$ \zeta_3^2= \left(\alpha +2 M \left(2 C R^3+R\right)-2 C R^2 \left(\alpha +R^2\right)+R^2 \sqrt{\frac{8 \alpha  M}{R^3}+1}\right)^2-4 \left(\alpha -2 M R+R^2\right)^2 $.\\

Thus, we conclude the matching of the interior and exterior spacetime across the boundary interface $r = R$. Note we have reverted to the canonical coordinates in the above calculations.

\end{itemize}

\subsection{Hydrostatic equilibrium}

Given the significance of the above results in the astrophysical context, it would be interesting if some comment regarding stability of the solution can be made. For this purpose we discuss the equilibrium stage that can be achieved by formulating the modified  TOV equation (\ref{tov}). Along these lines, a wide variety of  astrophysical solutions (including wormholes/compact stars) have been studied  (for a review, see, e.g., Refs. \cite{Rahaman:2013xoa,Rahaman:2014dpa,Jawad:2015uea,Maurya:2018kxg}). For compact star, the equilibrium picture can be discussed via, 
\begin{eqnarray}\label{mtov}
-\frac{dp}{dr}-\Phi'(\rho+p)=0, ~~~\text{or}~~~F_h+F_g =0, 
\end{eqnarray}
where the first term represents the hydrodynamic force ($F_h $) and the second term is gravitational force ($F_g$), respectively. The 
 Eq. (\ref{mtov}) predicts that the system is in equilibrium if the sum of different forces must be zero. As is evident in the plot of Fig. \ref{s1}, that the equilibrium of the forces is achieved for our chosen parametric values and confirms stability of the system.

\begin{figure*}
\centering
\includegraphics[width=0.6\linewidth]{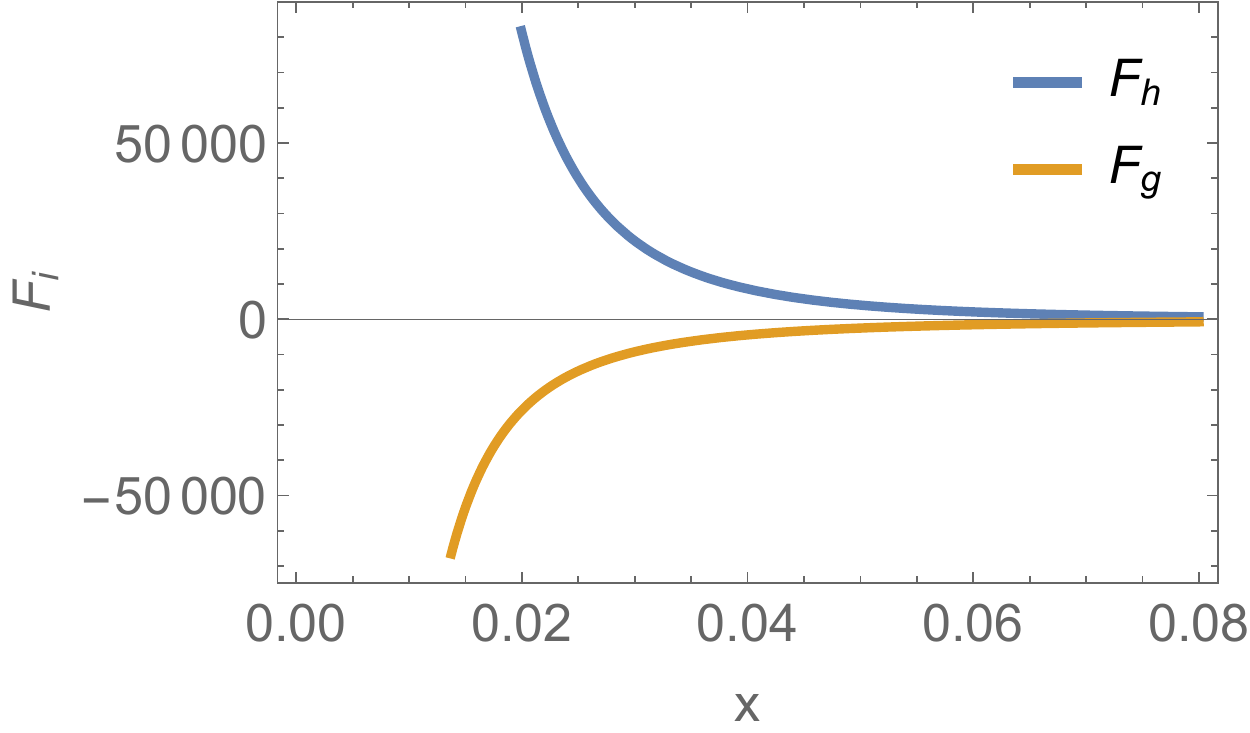}
\caption{Variation of the hydrodynamic and gravitational forces with respect to the radial coordinate $x$. } \label{s1}
\end{figure*}

\subsection{Stability under radial pulsations}

Models of perfect fluids are expected to be stable under radial pulsations. Historically, the first comprehensive treatment of stability was  by Chandrasekhar \cite{chandra} who developed a test for stability under perturbations of the pressure and energy density for a  linearised system of Einstein's equations. We emphasise that his scheme is only applicable to the Einstein's equations and not directly useful in our work with the injection of more complex higher curvature terms. In fact, it would be an interesting programme to emulate his work in this context but this is bound to be very complicated. The analysis boiled down to examining a boundary value problem of Sturm-Liouville type with the use of some trial functions  \cite{moustakidis}. Bardeen \textit{et al} soon thereafter devised a catalogue of methods to study the normal modes of radial pulsations of stellar models. A normal mode of radial pulsation for an equilibrium solution has the form $\delta r = \xi (r) \exp (i \sigma t)$ where $\xi$ is a trial function and the eigenvalue $\sigma$ is the frequency. We note that even though we start with a static distribution, if it undergoes some disturbance from the equilibrium position then it is expected that the configuration will oscillate and consequently be time dependent.  Additionally it is assumed, for this process that, the Lagrangian change of the pressure vanishes at the boundary and that the trial function is finite at the centre.  

It is worthwhile noting that studies of stability can be treacherous territory. Due warnings are contained in the work of Bardeen \textit{et al} \cite{bardeen}. For example, Knutsen \cite{knutsen}, applying method 2D of the Bardeen \textit{et al} \cite{bardeen} scheme,  came to the conclusion that the well known Buchdahl exact  solution is unstable with respect to radial oscillations. Contradictory results were obtained by  Negi \cite{negi} and Moustakidis \cite{moustakidis}. Kokkotas and Ruoff \cite{kokkotas} utilise a numerical technique and give two approaches to studying stability: the shooting method and the method of finite differences.    In the context of EGB gravity the stability of scalarized black holes has been considered by Silva \textit{et al} \cite{silva}.   In our case, since there is no worked out equivalent of the Chandrasekhar integral condition for EGB gravity, we shall analyse our exact solution under radial pulsations by perturbing the radial coordinate $x$ as $x + \epsilon x$, where $\epsilon$ is a very small quantity in relation to  the stellar radius, in the energy density, pressure and sound speed. It will be immediately apparent that for small values of $\epsilon$ the dynamical quantities converge to the original shapes.  We provide plots where $\epsilon$ is of the order $0.1, 0.01, 0.001$ in order to confirm that our solution is indeed stable under radial pulsations. This is in the same spirit as the well known Lyapunov stability of curves and follows the approach of Herrera \cite{herrera,abreu} in discussing the concept of cracking in anisotropic spheres. Of course, cracking does not apply to perfect fluids of the type we are studying here.  

Fig. 7, 8 and 9 depicts the energy density, pressure and sound speed  profiles under pulsations of the type $x + \epsilon x$ for three different $\epsilon$ values namely $0.1, 0.01, 0.001$ and each plot showing increments up to seven curves. It is clear that in all cases as $\epsilon$ decreases the perturbed curves tend toward the equilibrium curve $\epsilon = 0$. This indicates that these dynamical quantities are stable under radial pulsations. Moreover, in the case of the isotropic particle pressure, it is evident that even when the $\epsilon$ values are large $(0.1)$ the curves all tend to converge at the same stellar radius where the pressure vanishes. 

\begin{figure*}
\centering
\includegraphics[width=1\linewidth]{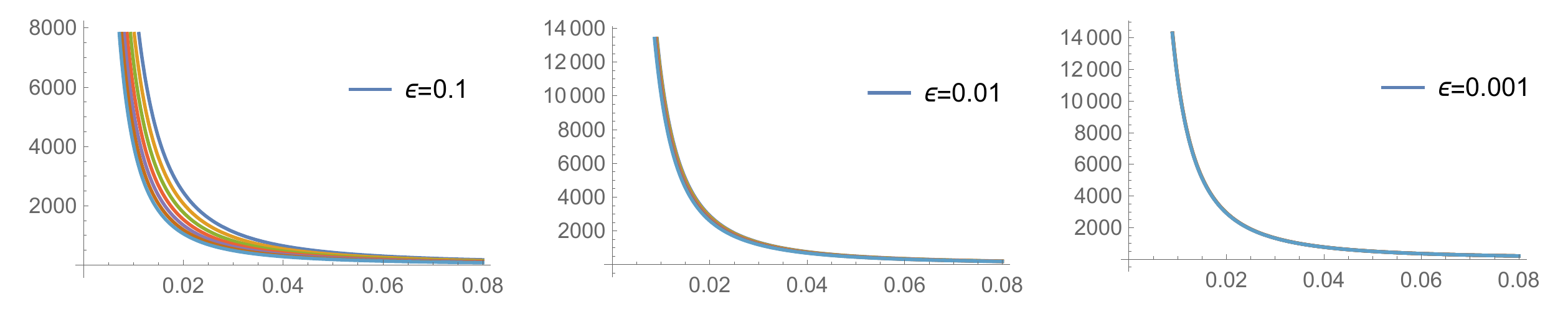}
\caption{Energy density versus radial pulsation $x + \epsilon x$ for $\epsilon$ of the order  $ 0.1, 0.01$ and $0.001$, respectively. The parameters are the same as given in Fig. \ref{fig:density}.  } \label{s1}
\end{figure*}

\begin{figure*}
\centering
\includegraphics[width=1\linewidth]{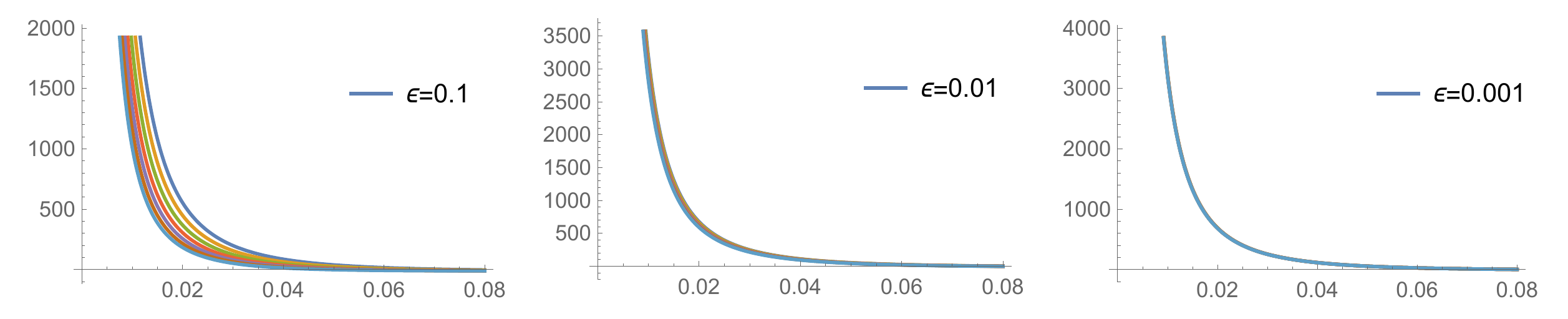}
\caption{Pressure versus radial pulsation $x + \epsilon x$ for $\epsilon$ of the order  $ 0.1, 0.01$ and $0.001$, respectively.   } \label{s1}
\end{figure*}

\begin{figure*}
\centering
\includegraphics[width=1\linewidth]{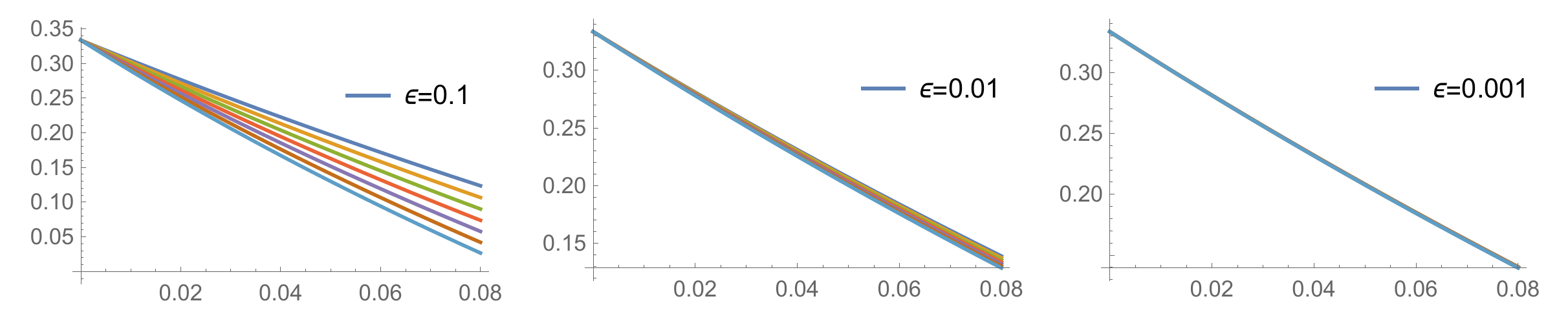}
\caption{Sound speed squared versus radial pulsation $x + \epsilon x$ for $\epsilon$ of the order  $ 0.1, 0.01$ and $0.001$, respectively.   } \label{s1}
\end{figure*}

%%%%%%%%%%%%%%%%%%%%%%%%%%%%%%%%%%%%%%%
\section{Additional exact Solutions}\label{sec5}

It is extremely difficult finding exact solutions to the field equations when higher curvature terms are present. Nevertheless, some success was achieved in locating a few classes of exact solutions which are listed below. However, we do not study these in detail. Despite several efforts we were unable to determine a suitable parameter space for these solutions such that all the physical requirements are satisfied. Accordingly we merely list these as exact solutions and in some cases we write out the dynamical quantities in detail where this is simple. 

\subsection{Constant space potential $Z = k$}

It is interesting to examine the case of a constant spatial potential $Z = k$, for some constant $k$ since it is known that this is a necessary and sufficient condition for isothermal behaviour in standard EGB gravity. By isothermal behaviour we mean that the density and pressure vary as $\frac{1}{r^{d-2}}$ where $d$ is the dimension of the spacetime. In standard Einstein gravity the isothermal property demands that the energy density and pressure go as $\frac{1}{r^2}$ as in the Newtonian gravity theory. In the current context the choice $Z = k$ results in the form 
\be
4kx^2 (\beta (1 -  k) +x) \ddot{y}+  4 k x \beta  (k-1) \dot{y}  -(k-1)  (\beta  (k-1)+x)y = 0, \label{110}
\ee
for the isotropy equation (\ref{100}). 
This selection gives 
\begin{eqnarray}
y &=& C_{12}F_1 \left(\left[-\frac{-3 \sqrt{2} \sqrt{4 k-2}+2 \sqrt{3} \sqrt{9 k+3}-6 \sqrt{k}}{12 \sqrt{k}},-\frac{\sqrt{2} \sqrt{4 k-2}+\frac{2}{3} \sqrt{3} \sqrt{9 k+3}-2 \sqrt{k}}{4 \sqrt{k}}\right], \right. \nonumber\\ 
&&\left.\left[-\frac{\sqrt{3} \sqrt{9 k+3}-3 \sqrt{k}}{3 \sqrt{k}}\right],\frac{x}{\beta  (k-1)}\right)x^{-\frac{\sqrt{3} \sqrt{9 k+3}-6 \sqrt{k}}{6 \sqrt{k}}} \nonumber\\
&&+ C_{22} F1\left(\left[\frac{3 \sqrt{2} \sqrt{4 k-2}+2 \sqrt{3} \sqrt{9 k+3}+6 \sqrt{k}}{12 \sqrt{k}},-\frac{\sqrt{2} \sqrt{4 k-2}-2 \sqrt{k}-\frac{2}{3} \sqrt{3} \sqrt{9 k+3}}{4 \sqrt{k}}\right], \right.\nonumber\\
&& \left.\left[\frac{\sqrt{3} \sqrt{9 k+3}+3 \sqrt{k}}{3 \sqrt{k}}\right],\frac{x}{\beta  (k-1)}\right)x^{\frac{\sqrt{3} \sqrt{9 k+3}+6 \sqrt{k}}{6 \sqrt{k}}}
\end{eqnarray}
in terms of hypergeometric functions and where $C_{11}$ and $C_{22}$ are integration constants. As it was not possible to find constant values $k$ for which $y$ could reduce to elementary functions it cannot be checked if isothermal behaviour results in 4D EGB as it does in standard EGB theory.

\subsection{$Z = x$}

The prescription $Z = x$ is a case of Kuchowicz's \cite{kuch} exact solution in Einstein theory where the radial potential was chosen in the form $x^{(b-1)}$ for some natural number $b$. The general solution was obtained for general $b$ but it was not observed that the $b = 2$ solution could be resolved as trigonometric functions. In our more complicated case, the $ b=2$ case   gives 
\begin{eqnarray}
y &=& c_1 e^{-\frac{1}{\sqrt{x}}} x-c_2 e^{\frac{1}{\sqrt{x}}} x, \\
8\pi \rho &=& -\frac{1}{2 x^2}-\frac{3}{2},\\
8\pi p &=& \frac{3}{2}+\frac{1}{2 x^2}\left(1 + 4 x+ \sqrt{x}\left(8c_1\left(c_1-c_2 e^{\frac{2}{\sqrt{x}}}\right)^{-1} -4\right) \right),
\end{eqnarray}
for $\beta = 1$. It was not possible to determine the solution for all $\beta$. Given the simplicity of the present solution 
we list the sound speed squared index and energy conditions
\begin{eqnarray}
\frac{dp}{d\rho} &=& -\frac{c_1^2 \left(2 x+3 \sqrt{x}+1\right)-2 c_2 c_1 e^{\frac{2}{\sqrt{x}}} (2 x-1)+c_2^2 e^{\frac{4}{\sqrt{x}}} \left(2 x-3 \sqrt{x}+1\right)}{\left(c_1-c_2 e^{\frac{2}{\sqrt{x}}}\right){}^2}, \\
8\pi (\rho - p) &=& -\frac{4 c_1}{x^{3/2} \left(c_1-c_2 e^{\frac{2}{\sqrt{x}}}\right)}+\frac{2}{x^{3/2}}-\frac{1}{x^2}-\frac{2}{x}-3,\\
8\pi (\rho + p) &=& \frac{2 \left(c_1 \left(\sqrt{x}+1\right)-c_2 e^{\frac{2}{\sqrt{x}}} \left(\sqrt{x}-1\right)\right)}{x^{3/2} \left(c_1-c_2 e^{\frac{2}{\sqrt{x}}}\right)},\\
8\pi (\rho + 3p) &=& \frac{12 c_1}{x^{3/2} \left(c_1-c_2 e^{\frac{2}{\sqrt{x}}}\right)}-\frac{6}{x^{3/2}}+\frac{1}{x^2}+\frac{6}{x}+3.
\end{eqnarray}
Finally  the Chandrasekhar adiabatic stability index is given by 
\begin{eqnarray}
\left(\frac{\rho+p}{p}\right)\frac{dp}{d\rho} &=& -\frac{4 \sqrt{x} \left(c_1 \left(\sqrt{x}+1\right)-c_2 e^{\frac{2}{\sqrt{x}}} \left(\sqrt{x}-1\right)\right){}^2 \left(c_1 \left(2 \sqrt{x}+1\right)-c_2 e^{\frac{2}{\sqrt{x}}} \left(2 \sqrt{x}-1\right)\right)}{\left(c_1-c_2 e^{\frac{2}{\sqrt{x}}}\right){}^2 \left(c_1 \left(3 x^2+4 x+4 \sqrt{x}+1\right)-c_2 e^{\frac{2}{\sqrt{x}}} \left(3 x^2+4 x-4 \sqrt{x}+1\right)\right)}. \nonumber \\
\end{eqnarray}
Despite its simplicity, it was not possible to find a solution which displayed pleasing physical properties.

\subsection{$y= \sqrt{1+x}$}

The stipulation $y = \sqrt{1+x}$ corresponds to the Schwarzschild interior temporal potential as can be seen from (\ref{100d}). It remains to see whether the the unique radial potential is indeed $Z = 1+x$. A major advantage of this choice of $y$ is that it is singularity free everywhere. The isotropy equation is solved to give   gives the general solution
\begin{eqnarray}
Z &=&\frac{1}{2 (2 x-1)}\left(\frac{x v }{k\beta}   
+   \frac{4 (\beta -1)^2 x k}{\beta v}  +  \frac{4 x^2}{\beta}+2 (x-1)\right), \nonumber \\ \label{110b}
\end{eqnarray}
where we have defined
\[
v =  \sqrt[3]{4\left(k^2 (x+1) (1-2 x)^2 -2 k^3 (\beta -1)^3  + k^2 (1-2x)\sqrt{  (x+1)  \left(4 x^3-3 x+1 -4 (\beta -1)^3 k\right)}\right)}.
\]
which is not exactly the radial potential of the constant density fluid. However, it is interesting to note that in the special case $\beta = 1$ 
 reduces to a much simpler exact solution that is more tractable. Setting $\beta = 1$ gives 
\begin{eqnarray} 
Z = 1 + x \hspace{1cm} {\mbox{or}} \hspace{1cm}  Z = 1 + x+\frac{c_1 x \sqrt[3]{x+1}}{\sqrt[3]{1-2 x}},
\end{eqnarray}
the first being the expected Schwarzschild potential and the second being a totally new solution. We briefly investigate the properties of the second solution. 
For the density and pressure we have 
\begin{eqnarray} 
8\pi \rho &=& \frac{1}{2} \left(\frac{c_1^{2} (x (6 x-1)-3)}{(2 x-1) \sqrt[3]{1-2 x}^2 \sqrt[3]{x+1}}-3\right),\\
8\pi p &=& \frac{1}{2} \left(\frac{c_1^2 (1-3 x) \sqrt[3]{x+1}^2}{(x+1) \sqrt[3]{1-2 x}^2}+3\right),
\end{eqnarray} 
and it is clear that the density is not constant but varies with the radius. It can also be shown that the pressure does vanish for a finite radius thus the model may be used to model compact stars. In addition we  list the sound speed squared index and energy conditions 
\begin{eqnarray} 
\frac{dp}{d\rho} &=& \frac{(2x-1)(x-1)}{2 x^2-x-5}, \\
8\pi (\rho - p) &=& \frac{c_1^2 \left(6 x^2-3 x-1\right)}{(2 x-1) \sqrt[3]{1-2 x}^2 \sqrt[3]{x+1}}-3, \\
8\pi (\rho + p) &=& \frac{2 c c_1^2 (x-1)}{(2 x-1) \sqrt[3]{1-2 x}^2 \sqrt[3]{x+1}}, \\
8\pi (\rho + 3p) &=& \frac{c_1^2 ((7-6 x) x-3)}{(2 x-1) \sqrt[3]{1-2 x}^2 \sqrt[3]{x+1}}+3.
\end{eqnarray}
Finally  the Chandrasekhar adiabatic stability index is evaluates to \begin{eqnarray} 
\left(\frac{\rho+p}{p}\right)\frac{dp}{d\rho} &=& -\frac{4 c_1^2 (x-1)^2}{(x (2 x-1)-5) \left(c_1^2 (3 x-1)-3 \sqrt[3]{1-2 x}^2 \sqrt[3]{x+1}\right)} .
\end{eqnarray} 
The presence of a singularity at $x = \frac{1}{2}$ is not necessarily a debilitating factor in this model since the pressure could vanish before this value of the radius. However, it can be seen that the central density is zero at the stellar centre which is not usually palatable since we expect a density profile that decreases monotonically from the centre of the distribution. Moreover, following extensive empirical testing, it was not possible to obtain suitable parameter values such that all the elementary conditions are satisfied.

\section{Summary and Discussion }\label{sec6}

We have considered the impact of the Gauss--Bonnet higher curvature invariants on stellar structure in the 4D framework with a rescaled coupling constant. Our interest was in this proposal's suitability in generating models of compact stars. After converting the master pressure isotropy equation to a form of an Abel equation we obtained an exact solution by prescribing the temporal potential to be proportional to the stellar radius. This prescription permitted the solving of the Abel equation and the complete model could then be constructed. The physical properties of the model were studied and shown to satisfy the usual requirements of positivity of density and pressure, existence of a surface of vanishing pressure, all the energy conditions being met, a subluminal sound speed and adiabatic stability in the sense of Chandrasekhar. It turns out that the higher curvature terms had the effect of correcting the deficiencies of the associated Einstein model. Finally, some more exact solutions were discovered and reported on and which merit closer scrutiny in the future.   This study has demonstrated that compact star models satisfying basic physical requirements are admissible in the 4D EGB framework.

\section*{Acknowledgments}

\end{document}